\documentstyle[12pt]{article}
\newcommand{\NP}{Nucl. Phys. }
\newcommand{\PR}{Phys. Rev. }
\newcommand{\PRL}{Phys. Rev. Lett. }
\newcommand{\PL}{Phys. Lett. }

\begin{document}
\baselineskip=20pt

\pagenumbering{arabic}

\vspace{1.0cm}
\begin{flushright}
LU-ITP 2003/015
\end{flushright}

\begin{center}
{\Large\sf Note on CKM matrix renormalization}\\[10pt]
\vspace{.5 cm}

{Yi Liao}
\vspace{1.0ex}

{\small Institut f\"ur Theoretische Physik, Universit\"at Leipzig,
\\
Augustusplatz 10/11, D-04109 Leipzig, Germany\\}

\vspace{2.0ex}

{\bf Abstract}

\end{center}

A simple inspection of the one loop quark self-energy suggests a prescription
of the CKM matrix renormalization in the standard model. It leads to a CKM
matrix counterterm which is gauge parameter independent and satisfies the
unitarity constraint, and renormalized physical amplitudes which are gauge
parameter independent and smooth in quark mass difference. We make a point
that caution should be practiced when interpreting the CKM martix counterterm
in terms of those of parameters in a given representation due
to rephasing effects from renormalization. We show how this can be done using
the degrees of freedom in the on-shell renormalization scheme.

\begin{flushleft}
% corresponding to: (PACS 2003 version)
% renormalization, quark and lepton masses and mixing


\end{flushleft}

\newpage
\section{Introduction}
The Cabibbo-Kobayashi-Maskawa (CKM) matrix appearing in the charged current
sector of the standard model (SM) arises from a mismatch in the transformations
of the up-type and down-type quark fields that bring them from weak gauge
eigenstates to mass eigenstates
\cite{ckm}.
As the matrix contains free physical parameters, it will generally be subject
to renormalization. Concerning renormalization, there is an essential
difference between these parameters and other physical parameters in SM.
The latter parameters include the masses of all physical particles and
the fine structure constant. There is a physically natural way to define or
renormalize them: the mass of a physical particle can be identified with the
real part of the pole of the corresponding field propagator; and the fine
structure constant can be defined in terms of the Thomson cross section due
to a theorem which states that the cross section of a soft photon scattering
against a massive particle approaches its classical result in the low energy
limit. In contrast, it does not make much sense to speak of gauge eigenstates
beyond the Lagrangian level. There is thus no physically preferred way to
define the CKM matrix at higher orders.

The necessity to renormalize the CKM matrix in order to obtain an ultraviolet (UV)
finite result for a physical amplitude was first analyzed by Marciano and
Sirlin for two generations \cite{marciano}. The case of three generations
was then studied by Denner and Sack \cite{denner} in the on-shell
renormalization scheme of SM \cite{aoki}.
A prescription was proposed for
the counterterm of the matrix, which is a combination of the quark
wave-function renormalization constants specified in the on-shell scheme.
Unfortunately, the counterterm so determined turns out to contain a UV finite
part that is gauge parameter dependent \cite{gambino}, which should be
avoided for physical parameters. An alternative prescription was then suggested
\cite{gambino,kniehl}, which is based on the quark wave-function renormalization
constants determined at zero momentum and shown to be independent of gauge
parameter. Such a prescription necessarily departs from the on-shell
renormalization scheme. To work exclusively in terms of on-shell
renormalization constants, the authors of Ref. \cite{barroso} proposed to
renormalize the matrix with respect to a reference theory in which no mixing
occurs. The program was further improved and elaborated upon in Ref.
\cite{diener}. Yet another approach \cite{yamada} was developed on the pinch
technique that is often used to tackle the problem of gauge
parameter dependence. From that point of view, the original prescription of
Denner and Sack may be reinterpreted as one of many possible ways to separate
out a counterterm for the CKM matrix that is gauge parameter independent, and
is thus acceptable.
All of these approaches thus differ only in a UV finite and gauge parameter
independent part in the counterterm for the CKM matrix and are understood as
renormalization scheme dependence in Ref. \cite{pilaftsis}. The issue has
also been investigated in relation to the wave-function renormalization for
unstable particles \cite{espriu,zhou}.

In this work we will show how a simple inspection of the one loop contribution
to quark self-energies suggests a way of splitting them: one part that is UV
divergent but gauge parameter independent, to be absorbed into the CKM matrix
counterterm, and the other that is UV finite but gauge parameter dependent, to
be put back to form renormalized physical amplitudes. The CKM matrix
counterterm so obtained shares with all proposals made so far the requisite
properties: gauge parameter independence, unitarity constraints and absorption
of the remaining UV divergence in physical amplitudes. It also enjoys a nice
feature that is incorporated in the prescriptions of Refs.
\cite{denner,barroso,diener,yamada}; namely, the renormalized physical amplitudes are smooth
when the up-type (or down-type) quark masses approach each other. The result
is similar to that of Ref. \cite{yamada} and differs in the UV finite and
gauge parameter independent terms, while avoiding the heavy machinery of the
pinch technique. We will also make a point that seems to have not been
emphasized in the literature. When one works in a specific representation of
the CKM matrix which is often convenient in practical calculations, caution
must be exercised in interpreting the CKM matrix counterterm in terms of its
rotation angles and CP phase. There is a relative rephasing, i.e., a change of
representations between the bare and renormalized CKM matrices due to
renormalization effects. This rephasing must be removed before one can write
down the counterterms for those angles and phase. We show how this can be
done using the degrees of freedom available in the on-shell renormalization
scheme.

\section{Renormalization of the CKM Matrix}
As the $Wud$-type vertex is the only available interaction term among physical
particles that involves the CKM matrix, it is natural to use it as a reference
to renormalize the CKM matrix. The one-loop renormalized amplitude for the
decay $W^+\to u_{\alpha}\bar{d}_i$ is \cite{denner}
%\begin{eqnarray}
%{\cal A}&=&
%\displaystyle
%-\frac{e}{\sqrt{2}s_W}\bar{u}_{\alpha}\rlap/\epsilon P_Lv_i
%\left[\left(\frac{1}{2}\sum_{\beta}\delta U^{L\star}_{\beta\alpha}V_{\beta i}
%\right.\right.
%\nonumber\\
%&&\displaystyle
%+\left.\left.\frac{1}{2}\sum_jV_{\alpha j}\delta D^L_{ji}+\delta V_{\alpha i}
%\right)+V_{\alpha i}(1+\delta C)\right]
%\nonumber\\
%&+&{\rm other~terms}.
%\label{amp}
%\end{eqnarray}
\begin{equation}
\begin{array}{rcl}
{\cal A}&=&\displaystyle
-\frac{e}{\sqrt{2}s_W}\bar{u}_{\alpha}\rlap/\epsilon P_Lv_i
\left[\left(\frac{1}{2}\sum_{\beta}\delta U^{L\star}_{\beta\alpha}V_{\beta i}
+\frac{1}{2}\sum_jV_{\alpha j}\delta D^L_{ji}+\delta V_{\alpha i}
\right)+V_{\alpha i}(1+\delta C)\right]\\
&+&{\rm other~terms}.
\end{array}
\label{amp}
\end{equation}
The up-type and down-type quarks are distinguished by the greek and italic
letters, respectively, so that the $(\alpha,i)$ entry of the CKM matrix is
denoted as $V_{\alpha i}$ with the counterterm $\delta V_{\alpha i}$.
$\epsilon$ is the polarization vector of the $W^+$
boson and $P_L=(1-\gamma_5)/2$, $P_R=(1+\gamma_5)/2$.
We shall work throughout in the on-shell renormalization scheme.
$U^{L,R}=1+\delta U^{L,R}$ and $D^{L,R}=1+\delta D^{L,R}$ are, respectively, the
wave-function renormalization constant matrices for the left- or right-handed
up-type and down-type quark fields. When distinction over up- and down-type
quarks is not necessary, we use $Z^{L,R}=1+\delta Z^{L,R}$ instead and the
italic letters for the flavors. In the above formula,
$\delta C=e^{-1}\delta e-s^{-1}_W\delta s_W+1/2~\delta Z_W+C_{1{\rm PI}}$.
Here $\delta e$, $\delta s_W$, and $Z_W=1+\delta Z_W$ are, respectively, the
counterterms for the electromagnetic coupling, the sine of the weak mixing
angle and the wave-function renormalization constant for the $W$ boson,
while $C_{1{\rm PI}}$ stands for the on-shell 1PI vertex contribution which
shares the same Lorentz structure as the tree level amplitude.
It is sufficient for us to know that without $\delta V_{\alpha i}$ the quantity
in the above square parentheses is gauge parameter independent but UV
divergent. The remaining UV divergences are expected to be cancelled by
$\delta V_{\alpha i}$, which
in turn must meet at least two more conditions. First, it must satisfy
$\delta V^{\dagger}V+V^{\dagger}\delta V=0$ to guarantee the unitarity of both
bare and renormalized matrices so that the number of independent physical
parameters is not changed by renormalization. Second, it does not introduce
new gauge parameter dependence. The other
terms not explicitly displayed in the above formula are the on-shell 1PI
vertex contributions of different Lorentz structures. These terms
are separately UV finite and gauge parameter independent as they must, and
are thus of no concern for our analysis.

Let us first review how wave-function renormalization constants and mass
counterterms are determined in the on-shell renormalization scheme in the
presence of mixing \cite{aoki}. We denote the renormalized mixing or
self-energy for the transition $j\to k$ as $\Gamma_{kj}(p)$. For the purpose
of determining the above mentioned counterterms, only the dispersive part
in $\Gamma_{kj}(p)$ is retained. This will always be implied in the following
discussion. The Hermiticity of the effective action
then demands that $\gamma_0\Gamma^{\dagger}\gamma_0=\Gamma$. Thus, it can be
parametrized as
%\begin{eqnarray}
%\Gamma_{kj}(p)&=&\rlap/p P_LF^L_{kj}(p^2)+\rlap/p P_RF^R_{kj}(p^2)\nonumber\\
%&+&P_LF^S_{kj}(p^2)+P_RF^{S\star}_{jk}(p^2),
%\label{gamma}
%\end{eqnarray}
\begin{equation}
\begin{array}{rcl}
\Gamma_{kj}(p)&=&\rlap/p P_LF^L_{kj}(p^2)+\rlap/p P_RF^R_{kj}(p^2)
+P_LF^S_{kj}(p^2)+P_RF^{S\star}_{jk}(p^2),
\end{array}
\label{gamma}
\end{equation}
with $F^{L\star}_{kj}=F^L_{jk}$, $F^{R\star}_{kj}=F^R_{jk}$.
The on-shell renormalization conditions that the renormalized mass is
identified with the zero of the real part of the self-energy, that no mixing
occurs between two particles when either of them is on-shell, and that the
residue of the diagonal propagator at the pole is unity, are equivalent to the
following equations,
\begin{eqnarray}
\displaystyle\Gamma_{kj}(p)u_j(p)|_{p^2\to m_j^2}&=&0,\nonumber\\
\displaystyle\frac{1}{\rlap/p-m_j}\Gamma_{jj}(p)u_j(p)|_{p^2\to m_j^2}&=&1.
\label{ren}
\end{eqnarray}
At one loop level,
$\Gamma_{kj}(p)=(\rlap/p-m_j)\delta_{kj}+\Gamma^{\rm loop}_{kj}(p)
+\Gamma^{\rm ct}_{kj}(p)$. Since the counterterm contribution
$\Gamma^{\rm ct}_{kj}(p)$ fulfills the Hermitian property separately, the
latter must also be respected by the loop contribution.
$\Gamma^{\rm loop}_{kj}(p)$ can be parametrized as in Eq. $(\ref{gamma})$
with $F$'s replaced by $\Sigma$'s. We then have
$\Sigma^{L\star}_{kj}=\Sigma^L_{jk}$, $\Sigma^{R\star}_{kj}=\Sigma^R_{jk}$.

The first condition in Eq. $(\ref{ren})$ yields for $j\ne k$,
%\begin{eqnarray}
%\displaystyle\delta Z^L_{kj}&=&\displaystyle\frac{2}{m^2_k-m^2_j}
%\left[m^2_j\Sigma^L_{kj}+m_jm_k\Sigma^R_{kj}\right.
%\nonumber\\
%&&+\left. m_k\Sigma^S_{kj}+m_j\Sigma^{S\star}_{jk}\right](m^2_j),
%\nonumber\\
%\displaystyle\delta Z^R_{kj}&=&\displaystyle\frac{2}{m^2_k-m^2_j}
%\left[m^2_j\Sigma^R_{kj}+m_jm_k\Sigma^L_{kj}\right.
%\nonumber\\
%&&+\left. m_j\Sigma^S_{kj}+m_k\Sigma^{S\star}_{jk}\right](m^2_j),
%\label{offz}
%\end{eqnarray}
\begin{equation}
\begin{array}{rcl}
\displaystyle\delta Z^L_{kj}&=&\displaystyle\frac{2}{m^2_k-m^2_j}
\left[m^2_j\Sigma^L_{kj}+m_jm_k\Sigma^R_{kj}
+m_k\Sigma^S_{kj}+m_j\Sigma^{S\star}_{jk}\right](m^2_j),\\
\displaystyle\delta Z^R_{kj}&=&\displaystyle\frac{2}{m^2_k-m^2_j}
\left[m^2_j\Sigma^R_{kj}+m_jm_k\Sigma^L_{kj}
+m_j\Sigma^S_{kj}+m_k\Sigma^{S\star}_{jk}\right](m^2_j),
\end{array}
\label{offz}
\end{equation}
and for $j=k$,
\begin{equation}
\begin{array}{rcl}
\delta m_j&=&\displaystyle\frac{1}{2}
\left[m_j(\Sigma^L_{jj}+\Sigma^R_{jj})
+\Sigma^S_{jj}+\Sigma^{S\star}_{jj}\right](m^2_j),\\
\delta Z^L_{jj}-\delta Z^R_{jj}&=&\displaystyle
\left[\Sigma^R_{jj}-\Sigma^L_{jj}
+\frac{\Sigma^S_{jj}-\Sigma^{S\star}_{jj}}{m_j}\right](m^2_j).
\end{array}
\end{equation}
%\begin{eqnarray}
%\delta m_j&=&\displaystyle\frac{1}{2}
%\left[m_j\left(\Sigma^L_{jj}+\Sigma^R_{jj}\right)
%+\Sigma^S_{jj}+\Sigma^{S\star}_{jj}\right](m^2_j),
%\nonumber\\
%\delta Z^L_{jj}-\delta Z^R_{jj}&=&\displaystyle
%\left[\Sigma^R_{jj}-\Sigma^L_{jj}
%+\frac{\Sigma^S_{jj}-\Sigma^{S\star}_{jj}}{m_j}\right](m^2_j).
%\end{eqnarray}
The above diagonal equations are also covered by the second condition in Eq.
$(\ref{ren})$ using the Hermitian property. In addition, the condition also
yields the following results:
%\begin{equation}
%\begin{array}{rl}
%&{\rm Re}~\delta Z^L_{jj}
%\nonumber\\
%=&\displaystyle
%-\Sigma^L_{jj}(m^2_j)
%\nonumber\\
%&\displaystyle
%-m_j\left[m_j(\Sigma^{L\prime}_{jj}+\Sigma^{R\prime}_{jj})
%+(\Sigma^{S\prime}_{jj}+\Sigma^{S\prime\star}_{jj})\right](m^2_j),
%\nonumber\\
%&{\rm Re}~\delta Z^R_{jj}
%\nonumber\\
%=&\displaystyle
%-\Sigma^R_{jj}(m^2_j)
%\nonumber\\
%&\displaystyle
%-m_j\left[m_j(\Sigma^{L\prime}_{jj}+\Sigma^{R\prime}_{jj})
%+(\Sigma^{S\prime}_{jj}+\Sigma^{S\prime\star}_{jj})\right](m^2_j),
%\end{array}
%\end{equation}
\begin{equation}
\begin{array}{rcl}
{\rm Re}~\delta Z^L_{jj}&=&\displaystyle
-\Sigma^L_{jj}(m^2_j)-m_j\left[m_j(\Sigma^{L\prime}_{jj}+\Sigma^{R\prime}_{jj})
+(\Sigma^{S\prime}_{jj}+\Sigma^{S\prime\star}_{jj})\right](m^2_j),\\
{\rm Re}~\delta Z^R_{jj}&=&\displaystyle
-\Sigma^R_{jj}(m^2_j)-m_j\left[m_j(\Sigma^{L\prime}_{jj}+\Sigma^{R\prime}_{jj})
+(\Sigma^{S\prime}_{jj}+\Sigma^{S\prime\star}_{jj})\right](m^2_j),
\end{array}
\end{equation}
with
$\displaystyle\Sigma^{L\prime}_{jj}(m^2_j)
=\left.\frac{\partial}{\partial p^2}\Sigma^L_{jj}(p^2)\right|_{p^2=m_j^2}$ etc.
The mass counterterms and the off-diagonal wave-function renormalization
constants are uniquely determined, while the diagonal ones are determined up to
a difference in imaginary parts for each $j$,
\begin{equation}
\begin{array}{rcl}
{\rm Im}~\delta Z^L_{jj}-{\rm Im}~\delta Z^R_{jj}&=&\displaystyle
\frac{2}{m_j}{\rm Im}~\Sigma^S_{jj}(m^2_j).
\end{array}
\end{equation}
When ${\rm Im}~\Sigma^S_{jj}(m^2_j)=0$, as is the case at one loop in SM, we can
choose arbitrarily a common imaginary part for $\delta Z^{L,R}_{jj}$ for each $j$.
This freedom is already contained in the on-shell renormalization scheme
\cite{aoki}: for a given set of $Z^{L,R}$, a common rephasing
$Z^{L,R}\to EZ^{L,R}$ with
$E={\rm diag}(e^{i\varphi_1},e^{i\varphi_2},\cdots)$ does not leave any trace in
the kinetic terms and the interaction terms in the neutral current sector which
involve quark fields of the same type. The off-diagonal $Z^{L,R}_{kj}$ starts
at $O(e^2)$ and can thus feel the arbitrariness only at two loop level, while
the diagonal $Z^{L,R}_{jj}$ starts at $O(1)$ and the arbitrariness shows up
already at one loop level. However, in the charged current sector
where both types of quarks participate, the rephasing does affect the appearance
of the CKM matrix. When we are supposed to renormalize in
a specific representation of the CKM matrix, this rephasing degree of freedom
should be considered together with the CKM matrix renormalization. We will
illustrate how this can be done later on.

A peculiar feature in the off-diagonal renormalization constants is that they
become singular as the masses of the two quarks under consideration approach
each other. If the degeneracy is exact, it must be protected by some symmetry
from renormalization effects so that the two quarks decouple from mixing with
other quarks of the same type. More interesting is the case when the mass
difference of two quarks is much smaller than their mass scale. In such a case,
there is no reason that the CKM matrix, as independent physical parameters,
must be trivial with respect to these closely lying quarks of the same type.
Although this does not occur phenomenologically in SM, it is theoretically
natural to expect that physical amplitudes like Eq. $(\ref{amp})$ should be
smooth in the mass difference. As the nonsmoothness is caused by the mixing,
it should in physical amplitudes be reabsorbed into the counterterm for the
mixing matrix. Now we show that this can be readily arranged.

The quark wave-function renormalization constants appearing in Eq.
$(\ref{amp})$ can be decomposed as follows:
%\begin{equation}
%\begin{array}{rl}
%&\displaystyle
%\frac{1}{2}\sum_{\beta}\delta U^{L\star}_{\beta\alpha}V_{\beta i}
%+\frac{1}{2}\sum_jV_{\alpha j}\delta D^L_{ji}
%\nonumber\\
%=&\displaystyle
%\frac{1}{4}\left[\sum_{\beta}\left(\delta U^{L\star}_{\beta\alpha}
%+\delta U^{L}_{\alpha\beta}\right)V_{\beta i}
%+\sum_jV_{\alpha j}\left(\delta D^L_{ji}
%+\delta D^{L\star}_{ij}\right)\right]
%\nonumber\\
%+&\displaystyle
%\frac{1}{4}\left[\sum_{\beta}\left(\delta U^{L\star}_{\beta\alpha}
%-\delta U^{L}_{\alpha\beta}\right)V_{\beta i}
%+\sum_jV_{\alpha j}\left(\delta D^L_{ji}
%-\delta D^{L\star}_{ij}\right)\right].
%\end{array}
%\label{decompose1}
%\end{equation}
\begin{equation}
\begin{array}{rl}
&\displaystyle
\frac{1}{2}\sum_{\beta}\delta U^{L\star}_{\beta\alpha}V_{\beta i}
+\frac{1}{2}\sum_jV_{\alpha j}\delta D^L_{ji}\\
=&\displaystyle
\frac{1}{4}\left[\sum_{\beta}\left(\delta U^{L\star}_{\beta\alpha}
+\delta U^{L}_{\alpha\beta}\right)V_{\beta i}
+\sum_jV_{\alpha j}\left(\delta D^L_{ji}
+\delta D^{L\star}_{ij}\right)\right]\\
+&\displaystyle
\frac{1}{4}\left[\sum_{\beta}\left(\delta U^{L\star}_{\beta\alpha}
-\delta U^{L}_{\alpha\beta}\right)V_{\beta i}
+\sum_jV_{\alpha j}\left(\delta D^L_{ji}
-\delta D^{L\star}_{ij}\right)\right].
\end{array}
\label{decompose1}
\end{equation}
Consider first the Hermitian combination of $\delta Z^L$. Using Eq.
$(\ref{offz})$ and the Hermitian property of $\Sigma^{L,R}$, we see that the
off-diagonal terms are smooth in quark mass difference; the diagonal term is
proportional to ${\rm Re}~\delta Z^L_{jj}$ and free of the arbitrariness
mentioned above. This part cannot enter into $\delta V_{\alpha i}$ since it
does not fulfil the unitarity constraint. In contrast, the anti-Hermitian
combination does fulfil the constraint and is not smooth in mass difference.
It should thus play a role in constructing $\delta V_{\alpha i}$. The diagonal
term of the combination is
\begin{equation}
\displaystyle+\frac{i}{2}V_{\alpha i}{\rm ~Im}\left[
-\delta U^L_{\alpha\alpha}+\delta D^L_{ii}\right],
\label{diag}
\end{equation}
which is not fixed in the on-shell scheme due to the arbitrariness. For the
off-diagonal terms, we use Eq. $(\ref{offz})$ and the explicit one loop
results listed in Ref. \cite{kniehl} for $\Sigma^{L,R,S}$ in general $R_{\xi}$
gauge. Note that at this order only the charged current loops can contribute.
Using the loop integrals $B_0,B_1$ defined there and $A(m)=m^2[B_0(0,m,m)+1]$
and making free use of unitarity of $V$,
the contribution in the up-type sector is arranged as follows:
%\begin{widetext}
%\begin{equation}
%\begin{array}{rl}
%&\displaystyle
%\frac{1}{4}\sum_{\beta\ne\alpha}\left(\delta U^{L\star}_{\beta\alpha}
%-\delta U^{L}_{\alpha\beta}\right)V_{\beta i}
%\nonumber\\
%=&\displaystyle
%-\frac{\alpha}{8\pi s^2_Wm_W^2}\sum_{\beta\ne\alpha}\sum_{j}
%V_{\beta i}V_{\alpha j}V^{\star}_{\beta j}
%\left\{\frac{1}{4}\left[(\xi_Wm^2_W+m^2_j-m^2_{\alpha})
%B_0(\alpha,j,\xi_W)-(\alpha\to\beta)\right]
%\right.
%\nonumber\\
%&\displaystyle\left.
%+\frac{1}{2}\frac{m^2_{\beta}+m^2_{\alpha}}{m^2_{\beta}-m^2_{\alpha}}A(j)
%+\frac{1}{2}\frac{1}{m^2_{\beta}-m^2_{\alpha}}\left[
%m^2_{\alpha}\left(m^2_WB_1(\alpha,j,W)
%-(m^2_W+m^2_j-m^2_{\alpha})B_1(\alpha,W,j)\right)
%+(\alpha\to\beta)\right]\right\},
%\end{array}
%\label{decompose2}
%\end{equation}
%\end{widetext}
\begin{equation}
\begin{array}{rl}
&\displaystyle
\frac{1}{4}\sum_{\beta\ne\alpha}\left(\delta U^{L\star}_{\beta\alpha}
-\delta U^{L}_{\alpha\beta}\right)V_{\beta i}\\
=&\displaystyle
-\frac{\alpha}{8\pi s^2_Wm_W^2}\sum_{\beta\ne\alpha}\sum_{j}
V_{\beta i}V_{\alpha j}V^{\star}_{\beta j}\\
\times&\displaystyle
\left\{\frac{1}{4}\left[(\xi_Wm^2_W+m^2_j-m^2_{\alpha})
B_0(\alpha,j,\xi_W)-(\alpha\to\beta)\right]
\right.%\\
%&\displaystyle
+\frac{1}{2}\frac{m^2_{\beta}+m^2_{\alpha}}{m^2_{\beta}-m^2_{\alpha}}A(j)\\
&\displaystyle\left.
+\frac{1}{2}\frac{1}{m^2_{\beta}-m^2_{\alpha}}\left[
m^2_{\alpha}\left(m^2_WB_1(\alpha,j,W)
-(m^2_W+m^2_j-m^2_{\alpha})B_1(\alpha,W,j)\right)
+(\alpha\to\beta)\right]\right\},
\end{array}
\label{decompose2}
\end{equation}
where we have used the abbreviations
\begin{eqnarray*}
A(j)&=&A(m_j),\\
B_1(\alpha,W,j)&=&B_1(m^2_{\alpha},m_W,m_j),\\
B_0(\alpha,j,\xi_W)&=&B_0(m^2_{\alpha},m_j,\sqrt{\xi_W}m_W),
\end{eqnarray*}
etc.
The second and third terms in the above are singular in mass difference,
UV divergent but $\xi_W$ independent, while the first one is smooth in mass
difference, UV finite upon summation over $j$ but contains a $\xi_W$-dependent
UV finite part. It is thus natural to absorb the second and third terms into
$\delta V_{\alpha i}$ and put the first term together with the Hermitian
combination back into Eq. $(\ref{amp})$ to form the renormalized amplitude.
Such a rearrangement meets all requisite conditions and also incorporates the
smoothness property into physical amplitudes. Including the down-type part, we
propose the following counterterm for the CKM matrix:
%\begin{widetext}
%\begin{equation}
%\begin{array}{rl}
%&\displaystyle
%\left(\frac{\alpha}{16\pi s^2_Wm_W^2}\right)^{-1}
%\tilde{\delta}V_{\alpha i}
%\nonumber\\
%=&\displaystyle
%\sum_{\beta\ne\alpha}\sum_{j}
%\frac{V_{\beta i}V_{\alpha j}V^{\star}_{\beta j}}
%{m^2_{\beta}-m^2_{\alpha}}\left\{
%(m^2_{\beta}+m^2_{\alpha})A(j)
%+\left[
%m^2_{\alpha}\left(m^2_WB_1(\alpha,j,W)
%-(m^2_W+m^2_j-m^2_{\alpha})B_1(\alpha,W,j)\right)
%+(\alpha\to\beta)\right]\right\}
%\nonumber\\
%+&\displaystyle
%\sum_{j\ne i}\sum_{\beta}
%\frac{V_{\beta i}V_{\alpha j}V^{\star}_{\beta j}}
%{m^2_j-m^2_i}\left\{
%(m^2_j+m^2_i)A(\beta)
%+\left[
%m^2_i\left(m^2_WB_1(i,\beta,W)
%-(m^2_W+m^2_{\beta}-m^2_i)B_1(i,W,\beta)\right)
%+(i\to j)\right]\right\}.
%\end{array}
%\label{delv1}
%\end{equation}
%\end{widetext}
\begin{equation}
\begin{array}{rl}
&\displaystyle
\left(\frac{\alpha}{16\pi s^2_Wm_W^2}\right)^{-1}
\tilde{\delta}V_{\alpha i}\\
=&\displaystyle
\sum_{\beta\ne\alpha}\sum_{j}
\frac{V_{\beta i}V_{\alpha j}V^{\star}_{\beta j}}
{m^2_{\beta}-m^2_{\alpha}}\left\{
(m^2_{\beta}+m^2_{\alpha})A(j)
\right.\\
&\displaystyle\left.+\left[
m^2_{\alpha}\left(m^2_WB_1(\alpha,j,W)
-(m^2_W+m^2_j-m^2_{\alpha})B_1(\alpha,W,j)\right)
+(\alpha\to\beta)\right]\right\}\\
+&\displaystyle
\sum_{j\ne i}\sum_{\beta}
\frac{V_{\beta i}V_{\alpha j}V^{\star}_{\beta j}}
{m^2_j-m^2_i}\left\{
(m^2_j+m^2_i)A(\beta)
\right.\\
&\displaystyle\left.+\left[
m^2_i\left(m^2_WB_1(i,\beta,W)
-(m^2_W+m^2_{\beta}-m^2_i)B_1(i,W,\beta)\right)
+(i\to j)\right]\right\}.
\end{array}
\label{delv1}
\end{equation}

Let us compare $\tilde{\delta}V$ with the prescriptions in Refs.
\cite{denner,yamada}. Denner and Sack proposed to absorb completely
into $\delta V$ the anti-Hermitian combination in Eq. $(\ref{decompose1})$
evaluated in the 't Hooft-Feynman gauge ($\xi_W=1$). This introduces UV
finite $\xi_W$ dependence into $\delta V$ and thus also the renormalized
amplitude \cite{gambino}, as shown here in the first term of Eq.
$(\ref{decompose2})$. Yamada applied the pinch technique to guide the splitting
of the quark wave-function renormalization constants. As the $\xi_W=1$
gauge serves as a reference point in the technique, the part split into the
renormalized amplitude vanishes in the $\xi_W=1$ gauge while the other
part split into $\delta V$ is the same as in the Denner-Sack prescription.
Thus the two prescriptions are essentally identical but now appear as an
acceptable construction in that context. The difference to Eq. $(\ref{delv1})$
is also clear: including in Eq. $(\ref{delv1})$ the first term of Eq.
$(\ref{decompose2})$ evaluated in the $\xi_W=1$ gauge (and its counterpart in the
down-type sector) goes back to their prescriptions. Such scheme dependence
seems unavoidable in CKM matrix renormalization. There seems to be no simple
relations to the prescriptions suggested in Refs.
\cite{gambino,barroso,diener} as they either employ subtraction at zero
momentum or make reference to a theory with no mixing.

In practice it is often convenient to work with a specific representation of the
CKM matrix although physical results are rephasing invariant and cannot depend
on which representation we use. When doing so, we must be careful in
interpreting $\delta V$ in terms of counterterms for rotation angles and CP
phase introduced in a given representation \cite{ft1}.
To see the point, we take a look at
the UV divergent terms in Eq. $(\ref{delv1})$ which are universal to all
prescriptions suggested so far (in $4-2\epsilon$ dimensions),
%\begin{widetext}
\begin{equation}
\displaystyle
\left(\frac{3}{2\epsilon}~\frac{\alpha}{16\pi s^2_Wm_W^2}\right)^{-1}
\tilde{\delta} V_{\alpha i}^{\rm div}
=\sum_{\beta\ne\alpha}\sum_{j}
V_{\beta i}V_{\alpha j}V^{\star}_{\beta j}
\frac{m^2_{\beta}+m^2_{\alpha}}{m^2_{\beta}-m^2_{\alpha}}m^2_j
+\sum_{j\ne i}\sum_{\beta}
V_{\beta i}V_{\alpha j}V^{\star}_{\beta j}
\frac{m^2_j+m^2_i}{m^2_j-m^2_i}m^2_{\beta}.
\end{equation}
%\end{widetext}
We use $\alpha,\beta,\gamma$ ($i,j,k$) to distinguish the three up-type
(down-type) quarks, and denote the rephasing invariant CP violating
parameter as
$J={\rm Im}[V_{\alpha j}V_{\beta i}V^{\star}_{\alpha i}V^{\star}_{\beta j}]$.
Then,
%\begin{widetext}
%\begin{equation}
%\left(\frac{3}{2\epsilon}~\frac{\alpha}{16\pi s^2_Wm_W^2}\right)^{-1}
%\frac{|V_{\alpha i}|^2}{J}
%{\rm Im}\left[\frac{\tilde{\delta} V_{\alpha i}^{\rm div}}{V_{\alpha i}}
%\right]
%=-2(m^2_{\beta}-m^2_{\gamma})(m^2_j-m^2_k)\left[
%\frac{m^2_{\alpha}}{(m^2_{\beta}-m^2_{\alpha})(m^2_{\gamma}-m^2_{\alpha})}
%+\frac{m^2_i}{(m^2_j-m^2_i)(m^2_k-m^2_i)}
%\right].
%\end{equation}
%\end{widetext}
\begin{equation}
\begin{array}{rl}
&\displaystyle
\left(\frac{3}{2\epsilon}~\frac{\alpha}{16\pi s^2_Wm_W^2}\right)^{-1}
\frac{|V_{\alpha i}|^2}{J}
{\rm Im}\left[\frac{\tilde{\delta} V_{\alpha i}^{\rm div}}{V_{\alpha i}}
\right]\\
=&\displaystyle
-2(m^2_{\beta}-m^2_{\gamma})(m^2_j-m^2_k)\left[
\frac{m^2_{\alpha}}{(m^2_{\beta}-m^2_{\alpha})(m^2_{\gamma}-m^2_{\alpha})}
+\frac{m^2_i}{(m^2_j-m^2_i)(m^2_k-m^2_i)}
\right].
\end{array}
\end{equation}
The above indicates clearly that an imaginary part will be induced for the
element $\tilde{\delta} V_{\alpha i}$ even if we start with a representation
in which the element $V_{\alpha i}$ is real. Namely, there is a rephasing
effect due to renormalization which brings the CKM matrix from one
representation before renormalization to another after. On the other
hand, it is natural to require that a real bare element should have a real
counterterm. Fortunately, this can be accommodated by employing the degrees
of freedom in the on-shell renormalization scheme.

So far we have been focusing on the off-diagonal part in the anti-Hermitian
combination of wave-function renormalization constants, but said nothing about
the diagonal part shown in Eq. $(\ref{diag})$. Using field rephasing we can
make five out of nine elements in the CKM matrix real while the remaining four
will contain a representation dependent CP phase. This suggests the following
modification to the CKM counterterm,
\begin{equation}
\displaystyle\delta V_{\alpha i}
=\tilde{\delta}V_{\alpha i}
+\frac{i}{2}V_{\alpha i}{\rm ~Im}\left[
-\delta U^L_{\alpha\alpha}+\delta D^L_{ii}\right],
\end{equation}
which can be used to arrange that the bare and renormalized matrices are in the
same representation. For example, in the orginal Kobayashi-Maskawa
parametrization \cite{ckm}, the elements in the first row and column
are real. We adjust the second terms in the above equation for
$(\alpha,i)=(1,1),(1,2),(1,3),(2,1),(3,1)$ to absorb completely the imaginary
part contained in $\tilde{\delta}V_{\alpha i}$'s so that the corresponding
$\delta V_{\alpha i}$'s are real. This also fixes uniquely the second terms
for the other four complex elements. The final result can be cast in a compact
form applicable to all entries in the KM parametrization,
\begin{equation}
\displaystyle\frac{\delta V_{\alpha i}}{V_{\alpha i}}
=\frac{\tilde{\delta}V_{\alpha i}}{V_{\alpha i}}
-i\left[\frac{{\rm Im}~\tilde{\delta}V_{\alpha 1}}{V_{\alpha 1}}
+\frac{{\rm Im}~\tilde{\delta}V_{1i}}{V_{1i}}
-\frac{{\rm Im}~\tilde{\delta}V_{11}}{V_{11}}\right].
\end{equation}
Note that all nice features of $\tilde{\delta}V_{\alpha i}$
carry over to $\delta V_{\alpha i}$ since the above modification amounts to a
$\xi_W$ independent rephasing to the matrix and a reshuffling of terms in the
renormalized amplitude that are cancelled anyway. But $\delta V_{\alpha i}$
now allows for an interpretation in terms of rotation angles and CP phase.

\section{Conclusion}
We have shown in this work that a simple inspection of the quark self-energies
suggests a prescription for the CKM matrix renormalization. The obtained matrix
counterterm satisfies the unitarity constraint and is gauge parameter
independent. It also leads to a physical amplitude which is UV finite, gauge
parameter independent, and smooth in quark mass difference. We further improve
it using the freedom in the on-shell renormalization scheme so that the
counterterm also allows for an interpretation in terms of those of parameters
in any given representation of the CKM matrix.

\vspace{0.5cm}
\noindent
{\bf Acknowledgements}

%\begin{acknowledgments}
I would like to thank K. Sibold for reading the paper carefully.
%\end{acknowledgments}

% Create the reference section using BibTeX:
%\bibliography{basename of .bib file}

\begin{thebibliography}{30}
\bibitem{ft1}It has been proposed in Ref. \cite{diener} to express
directly the counterterm for the CKM matrix in terms of those of rotation
angles and CP phase in a given representation. However, as explained here,
this is not possible without removing a rephasing beforehand.
\end{thebibliography}
% I use the thebibliography environment instead

\end{document}